\documentstyle[prl,aps,epsf]{revtex}
\begin{document}
\draft
\twocolumn[\hsize\textwidth\columnwidth\hsize\csname@twocolumnfalse\endcsname

\title{The Effect of Splayed Pins on Vortex Creep and Critical Currents}
\author{C. J. Olson, R. T. Scalettar, G. T. Zim\'{a}nyi}
\address{Department of Physics, University of California, 
Davis, California 95616}
\date{\today}
\maketitle

\begin{abstract}
We study the effects of splayed columnar pins
on the vortex motion using realistic London Langevin simulations.
At low currents vortex creep is strongly suppressed,
whereas the critical current $j_c$ is enhanced only moderately.
Splaying the pins 
generates an increasing energy barrier against
vortex hopping, and leads to the forced entanglement of vortices, 
both of which suppress creep efficiently. 
On the other hand splaying enhances kink nucleation
and introduces intersecting pins, which cut off the energy barriers.
Thus the $j_c$ enhancement is strongly parameter sensitive. 
We also characterize the angle dependence of $j_c$,
and the effect of different splaying geometries.
\end{abstract}
\pacs{PACS numbers: 74.60.Ge,74.60.Jg}
\vskip2pc]

The enhancement of critical currents in superconductors through irradiation
by heavy ions is well established \cite{Civale1,Civale2}.  
The ions create extended columnar
defects that localize individual vortices much more effectively than 
naturally occurring point defects.  
It was suggested by T. Hwa {\it et al.} \cite{Hwa3} that pinning 
could be further improved by splaying these columnar defects.
In recent experiments 
\cite{Elbaum4,Kwok5,Hardy6,Elbaum7,Herbsommer8},
splayed pinning has been created directly
through multiple-step irradiation of the sample, or
indirectly by irradiation through thin foils.
Up to an order of magnitude enhancement of the critical current relative 
to parallel columnar pins has been reported \cite{Elbaum4}.

Columnar pins which are aligned with the field direction
allow a flux line to sit in a potential minimum in every layer without
having to pay any elastic energy.  There are no competing tendencies 
which favor point over columnar pinning, and hence there is
a clear expectation that columnar pinning will greatly enhance
the critical current,
as is universally born out experimentally. 

That splay should further enhance $j_c$ seems much less compelling.
It has been argued \cite{Hwa3} that a vortex hopping between
two splayed pins experiences a linearly increasing energy
cost, whereas once a hop
originates between columnar pins there is no further cost for 
completion of the jump.  This leads to a suppression of motion for
splayed pins.  
An additional mechanism for enhancement of $j_c$ that we suggest 
is an increase in collective pinning effects due to greater vortex
entanglement in the presence of
splayed pins.
This second mechanism may be more effective than the first since, 
although the distance between 
two splayed defects increases as one moves
away from their point of closest approach, the separation
between the pins may be very small at this closest point.
Thus one could argue that splaying leads to an {\it increased} nucleation
of hopping.  Furthermore, the increasing energy cost
which exists for two splayed pins can be cut off by the additional pins
that are present.
 
Set against the suggestions that splay enhances pinning is also the fact that
splaying undermines the
key feature which made columnar pinning so effective in the first place.
Vortices must elongate in order to take advantage of the pins,
though this wandering is not random, as for point pinning.
Even accepting the increasing energy cost for hopping and entanglement
arguments, it is not obvious that they
will dominate over the very 
special topological effectiveness of columnar pinning.
The uncertain consequences of this
set of competing effects is reflected experimentally in
the fact that the enhancement of pinning by splay is 
sensitive to the details of the splay, including whether the
angular distribution is Gaussian or bimodal, the splay angle, 
and the material used.

In spite of the importance of splay for critical current enhancement,
no realistic numerical simulations of vortices interacting with
splayed pinning have been reported up until this time.
In this paper we present the first analysis of the effects of
splayed pinning using overdamped molecular dynamics simulations.
By examining the distribution of lengths of pinned vortex segments between
kinks, we explicitly verify the conjecture
that the energy cost for vortex motion increases
for intermediate length segments in the presence of splay, 
as a two pin argument
suggests.  We refine this picture in a crucial way by demonstrating how this
growth is cut off by intersections with additional pins.
This measure at the same time shows an enhancement of
short length segments (nucleation events).
By comparing splay which is in the plane of the Lorentz force 
with splay that is orthogonal to it,
we can also address the effect of entanglement on pinning.
We find that the combination of entanglement and confinement
can lead to a suppression of creep, but only in a somewhat limited
parameter range.
Finally, we emphasize the need to distinguish suppression of vortex creep
and suppression of the critical current when determining the efficacy
of different pinning configurations.

We conduct overdamped molecular dynamics simulations using a 
London-Langevin model.  We refer the reader to \cite{vanOtterlo9} 
for details.  The key feature of the approach is the incorporation of
forms for the vortex-vortex interaction, elastic bending,
and thermal Langevin forces, which use experimental values for the 
coherence and penetration lengths, and anisotropy parameter, and hence
allows us to simulate the system without a host of adjustable parameters.
In \cite{vanOtterlo9} we verified that our choices accurately reproduce
the experimental phase diagram.  
The current simulations are performed at $T=77K$, 
inside the glassy phase as evidenced by the creep-like E-J curves below.
The samples are of size $106 \lambda \times
106 \lambda$, containing 49 vortices extending through 80 layers.

The pinning potential representing irradiated defects 
is modeled by short-range attractive parabolic wells of radius
$r_p = 0.5\lambda$ which are spatially correlated on neighboring
layers to form columnar pins.
The wells in different layers belonging to a single column have the
same pinning energy $U_p^i$, with $U_p^i$, 
selected from a Gaussian distribution with mean $U_p=0.08$ and standard
deviation $\sigma_p = 0.012$.
With this choice the vortex depinning transition falls near
$j/j_0 = 0.08$ 
(where $j_0$ is the BCS depairing current), 
and the magnitude of the pinning force is on average
of the same order as the elastic and Lorentz forces.
We consider several different pinning geometries: parallel columnar
pins, in which the columns are aligned with the $z$ axis; 
transverse bimodal splay, in which each pin is tilted at angles $\pm 
\theta$ from the $z$ axis in the plane transverse to the direction of vortex 
motion; longitudinal bimodal splay, in which the tilt plane is spanned 
instead by the direction of vortex motion and the $z$ axis; 
Gaussian splay, in which $\theta$ is selected from a Gaussian
distribution centered about zero and the tilt direction is uniformly 
distributed in the $x$--$y$ plane; 
and point pins, 
which are not 
spatially correlated between layers.

\begin{figure}
\center{
\epsfxsize=3.5in
\epsfbox{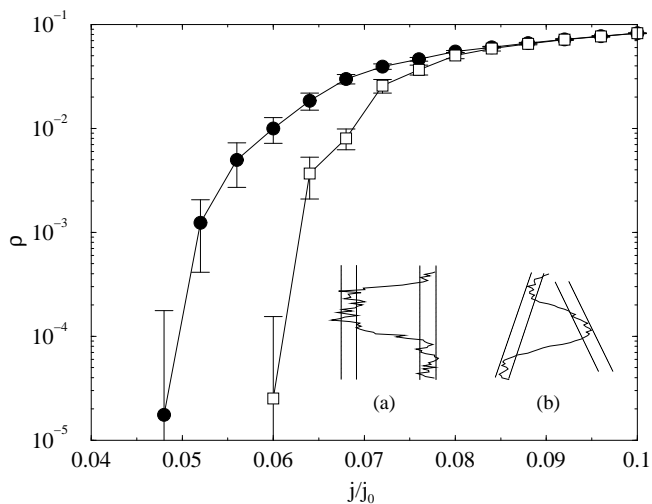}}
\caption{Resistivity $\rho$ as a function of applied current 
$j/j_{0}$ for columnar (filled circles) and transverse bimodal
splayed (open squares) defects, in samples containing 49 vortices, 
225 pins, and 80 layers. The splay angle $\theta =  \pm 5.7^{\circ}$.
A clear reduction of the resistivity is seen for splayed pinning
at small applied currents.
Inset:  
Simulation images of individual vortices forming double kinks
(a): between two columnar pins,  (b): between two splayed pins. 
}
\label{fig1}
\end{figure}

We first study single vortex phenomena 
by performing simulations in which
the number of vortices is less than the number of pins, $N_v < N_p$. 
In Fig.~\ref{fig1}, we compare two samples with an equal
density of columnar pins.
Here the dimensionless resistivity, $\rho = (E/j)/\rho_{BS}$
is defined in units of $\rho_{BS}$, the Bardeen-Stephen resistivity.
$E$ is the electric field and $j$ is the current density.
In the first sample the pins are aligned parallel
to the $z$ axis, and in the second the pins are splayed at
$\theta = \pm 5.7^{\circ}$ from the $z$ axis, transverse to the 
direction of vortex motion. We find {\it a strong suppression 
of the creep of vortices by splay in the low
current regime.} While the dynamic range is limited, when a creep
type exponential fit, $E \sim {\rm exp} (-1/j^{\mu})$, is performed, 
it is consistent with an increased value of $\mu$.

The effect of splaying decreases with increasing current.
The critical current $j_c$ is typically defined via a threshold criterion
$\rho(j_c) = \rho^{\rm t}$. Choosing low thresholds we clearly observe 
an {\it enhancement of the critical current}. For example, using 
$\rho^{\rm t} = 10^{-4}$, $j_c$ is enhanced by 20$\%$. 
At higher thresholds this enhancement is reduced, however,
finally disappearing for $\rho^{\rm t} > 0.05$.

Now we analyze the physical mechanisms at work in the presence of splay. 
Single vortex phenomena dominate in the dilute vortex limit, considered here.
At low applied currents, $j/j_0 < 0.1$, 
vortices move between pins by
thermally activated double kinks 
\cite{Blatter10,Yeshurun11,Geshkenbein12,Nelson13}.
For parallel columnar pins, extending an already formed double kink
does not cost extra energy [Fig.~\ref{fig1}(a), simulation image].
For splayed pins with a bimodal distribution, 
double kinks between pairs of pins tilted in {\it opposite} directions
[Fig.~\ref{fig1}(b), simulation image] 
experience an {\it increasing, or confining energy barrier}, 
since after the nucleation of the kink 
the unpinned vortex segment must keep growing longer in the high 
energy region {\it between} the pins. 
On the other hand, pairs of pins tilted in the {\it same} direction
with respect to the $z$ axis are twice as far apart on average than in the 
columnar case, so the nucleation of double kinks bridging 
parallel pins must span twice as long a distance, 
and hence costs more energy.

We extract the energy barrier against the spreading of kinks
by measuring the distribution function, $P(l_k)$, of the lengths $l_k$ 
of the vortex segments between kinks. If the kink energy does not 
depend on its length, as is expected for columnar pins, $P(l_k)$
should be roughly uniform for pins of equal depth, and exhibit slow 
decay with length for pins with differing depths.  
If the kink energy grows linearly
with length, as expected for splayed pins, $P(l_k)$
should fall off at large $l_k$ exponentially.

\begin{figure}
\center{
\epsfxsize=3.5in
\epsfbox{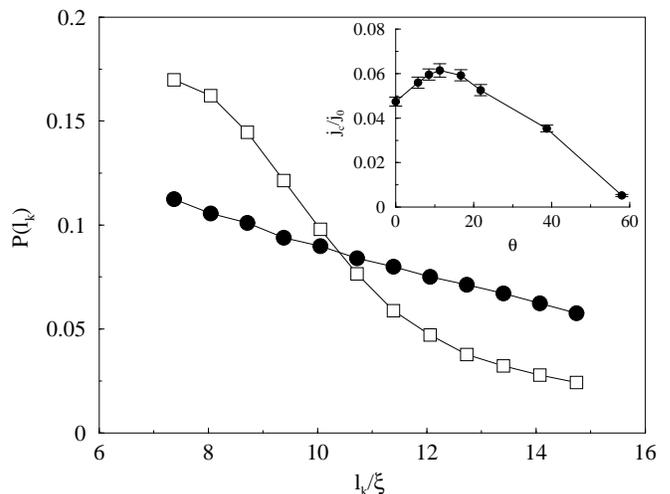}}
\caption{Distribution of the lengths of vortex segments between kinks, 
$P(l_k)$, for samples with columnar (filled circles) and splayed 
(open squares) pinning, with the same parameters shown in Fig.~\ref{fig1}
at $j/j_0 = 0.065$ (in the vortex creep regime).
For the columnar pinning, $P(l_k)$ depends weakly on length,
but $P(l_k)$ for splayed pinning falls off more rapidly,
indicating that the kink energy
grows with length $l_k$.  
Inset:
Critical current $j_{c}$ 
as a function of the angle $\pm \theta$
between the $z$ axis and the splayed pins 
for the system of Fig. 1.
The initial increase of $j_{c}$ with $\theta$ stops above 
$\theta \sim 10^{\circ}$
when the pins become so tilted that the vortices no longer accommodate
to them.
}
\label{fig2}
\end{figure}

In Fig.~\ref{fig2} we show $P(l_k)$ for $j/j_0=0.065$.
The columnar distribution is rather uniform,
whereas the splayed one exhibits a significant
enhancement at small $l_k$ relative to the columnar case, 
followed by a rapid fall in the intermediate region, and a slow 
decay at the large $l_k$ regime.
The enhancement at small $l_k$'s has a simple explanation:
because of the splaying, the minimal distance between the pins
is much smaller than for the columnar case. Therefore 
the {\it nucleation} of the double kinks which takes place here,
and the formation of short segments, is much enhanced.
The subsequent fall of $P(l_k)$ at intermediate $l_k$
is consistent with an exponential, supporting the picture of a linearly
increasing potential barrier. At large values $l_k \gtrsim 13\xi$, however, 
the decay of $P(l_k)$ is slowed down, 
and $P(l_k)$ tracks the columnar distribution. 
This can be attributed to the interference of additional pins.
The high energy segment between two tilted pins 
increases only until one of the
kinks reaches a {\it third} pin intersecting 
the second pin onto which the vortex is hopping.
In the present bimodal distribution this 
third pin is parallel to the first one.  
Thus, as the kinks slide further, the length of the high energy
segment remains unchanged, in complete analogy to the columnar case.

To consider many-vortex phenomena,
we move to the high density limit, $N_v > N_p$.
For $N_v = 2 N_p$, half of the vortices are directly pinned,
and the rest are pinned only by the repulsion of their pinned neighbors.
At low driving currents, only the latter, interstitial vortices move.
The flowing interstitial vortices are roughly aligned with the applied
magnetic field, whereas the pinned vortices are tilted. 
Thus the interaction between these two types of vortices results in
their entanglement.

\begin{figure}
\center{
\epsfxsize=3.5in
\epsfbox{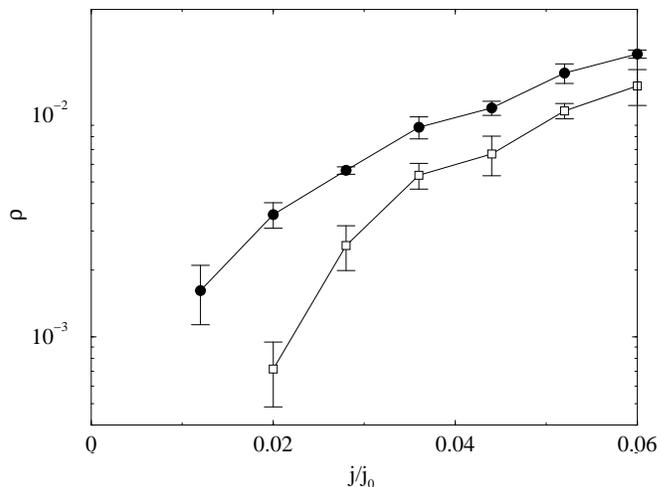}}
\caption{
Resistivity $\rho$ versus driving current $j/j_0$ for 
a sample containing 49 vortices and 25 pins.  Filled circles:
parallel columnar pinning; open squares: transverse bimodal
splayed pinning.
}
\label{fig3}
\end{figure}

Fig.~\ref{fig3} shows the result of our simulations.
First, the critical current is much reduced compared to the
case $N_v < N_p$, due to the fact that the interstitial vortices depin
much more easily than their pinned neighbors. 
Second, when $j_c$ is defined using the same threshold resistivity as before, 
$\rho^{\rm t} = 10^{-4}$, we observe a factor of 2 enhancement in
$j_c$ by splay, compared to the 20$\%$ seen in Fig.~\ref{fig1}. 
This strongly suggests that the forced entanglement of vortices
is capable of impressive additional enhancements of the critical current.

Here we are in the position to check whether the physics
of splayed pins is ``adiabatically connected" to the columnar case.
It is expected that either a single vortex is still sticking
to a single pin, after it has been splayed, 
or that several vortices will pin to a column in such
a way that columns are either (nearly) fully occupied, or (nearly)
completely empty, forming vortex-like ``quasi lines" \cite{Hwa3}. 
We tested these expectations by determining the distribution
function of percentage-wise occupation of the splayed pins.
Remarkably, this distribution function shows sharp peaks around
zero and 100$\%$ occupancy, with very suppressed values in between,
verifying the expectations. At small angles this is mostly due to 
vortices still sticking to a single pin.

We close this section by mentioning that the enhancement of $j_c$
occurred only for a rather limited range of the model's parameters;
for large regions of the parameter space we found either minimal effects,
or the reduction of $j_c$ upon splaying as the magnetic field was
further increased. This shows the powerful
influence of the enhanced kink nucleation and cutoff of the confining
potential by third pins.

Experimentally up to tenfold $j_c$ enhancements were reported 
\cite{Elbaum4,Hardy6}, exceeding but comparable to our results.
This enhancement is strongly dependent on the magnetic
field and temperature, however, often diminishing to small values,
or even turning into a reduction instead. This is consistent with
our finding of the importance of nucleation and intersecting pin effects.

Next we return to samples with $N_v < N_p$, and
explore the dependence of $j_c$ on the angle of splaying 
$\theta$.  The inset of Fig.~\ref{fig2} displays
$j_c(\theta)$. For small angles $\theta < 10^{\circ}$, $j_{c}$ increases 
due to the increased energy barrier to kink spreading with increasing
angle. Around $\theta \sim 10^{\circ}$ $j_{c}$ exhibits a maximum. 
It decreases smoothly for larger angles, as the vortices cease
to accommodate to the pins in order to stay aligned with the magnetic field
\cite{Blatter10,Nelson13}. This nonmonotonic behavior of $j_c$ is
consistent with experiments \cite{Elbaum4}.

Experimentalists have investigated several different pinning configurations 
\cite{Elbaum4,Kwok5,Hardy6,Elbaum7,Herbsommer8}. 
To make contact with these studies, 
we have measured the resistivity of samples with
{\it equal} numbers of pinning elements placed in the following
arrangements: uncorrelated point-like pinning, 
Gaussian splayed pinning, columnar pinning,
as well as longitudinal and transverse bimodal splayed pinning.

\begin{figure}
\center{
\epsfxsize=3.5in
\epsfbox{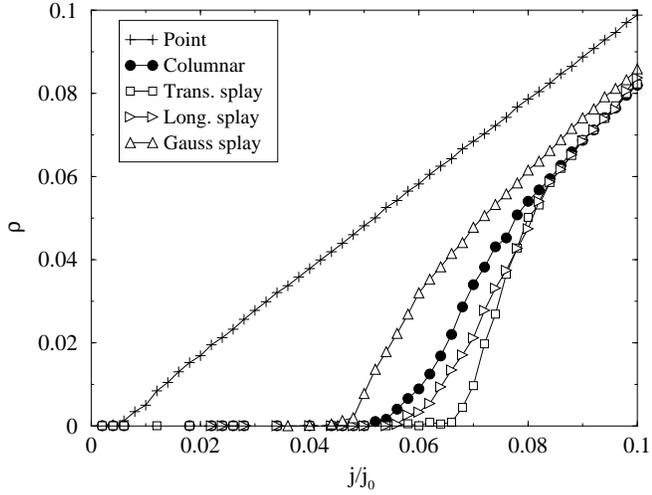}}
\caption{
Resistivity $\rho$ versus driving current $j/j_0$ for several
different pinning geometries: point pinning (plus signs), 
parallel columnar pinning (filled circles),
transverse bimodal (open squares) and longitudinal bimodal 
(open right triangles) splay pinning at $\theta = \pm 5^{\circ}$,
and Gaussian splay pinning 
with $\sigma_\theta = 5^{\circ}$ (open up triangles).
All correlated
pinning produces lower resistivities than point pinning.
Transverse bimodal splay produces a lower $\rho$ than columnar
pinning, but Gaussian splay produces a higher $\rho$.
}
\label{fig4}
\end{figure}

As shown in Fig.~\ref{fig4}, we find the highest resistivity for 
point pins, lower for columnar, and the lowest for bimodal transverse splay.
We find that the Gaussian splay produces an enhancement of the 
creep relative to columnar pins, which is consistent with experiments
\cite{Elbaum4,Elbaum7}.
Among the bimodal splay configurations,
transverse splay suppresses creep more than longitudinal splay,
again in agreement with experiments \cite{Kwok5}.
Transverse splay is more effective than longitudinal, because
it forces the entanglement of vortices more effectively.
Also, longitudinal splay brings the pins closer in the direction
of vortex motion, thus helping the nucleation of the double kinks.

Finally we return to the current dependence.
As the applied current increases to values approaching the depinning
regime, the vortices are pinned less
effectively and spend more time between pins. 
The vortex motion is no longer dominated by kinks, weakening the 
single vortex arguments. Also, forced entanglement is no longer effective
at preventing vortex motion near the depinning regime, since when a 
vortex begins to move, it is increasingly likely that
its forward neighbor is already moving, or that
the push from behind is sufficient to start its motion.
Thus instead of a forced entanglement, both vortices move together.
For all these reasons, the creep suppression by splay decreases
with increasing currents, and completely disappears in 
the depinning regime.

In conclusion, we have used realistic London Langevin
simulations to study the motion of vortices in the presence
of splayed columnar defects. 
We found that splaying introduces a confining potential against vortex
hopping. It also enhances kink nucleation and introduces third 
pins, however, cutting off this linear potential. We also established
the importance of forced entanglement.
Finally we analyzed the angle dependence of the critical current, and
compared different splaying geometries. Several of our results compare 
favorably to experiments.

We thank G. Crabtree, N. Gr{\o}nbech-Jensen, T. Hwa, L. Krusin-Elbaum, 
W. Kwok, P. Le Doussal, D. Nelson, and V. Vinokur for useful discussions.
Funding was provided by NSF-DMR-028535, and the CLC and CULAR grants,
administered by the University of California.

\end{document}